\documentclass{PoS}

\def\bea{\begin{eqnarray}}
\def\eea{\end{eqnarray}}
\def\pbpnull{\langle\bar{\psi_0}\psi_0\rangle}
\def\pbp{\langle\bar{\psi}\psi\rangle}

\title{QCD thermodynamics with Wilson fermions}

\ShortTitle{QCD thermodynamics with Wilson fermions}

\author{
Szabolcs Bors\'anyi$^{a}$,$\;$ 
Zolt\'an Fodor$^{abc}$,$\;$ 
Christian Hoelbling$^{a}$,$\;$ 
S\'andor D. Katz$^{c}$,$\;$ 
Stefan Krieg$^{ab}$,$\;$ 
\speaker{D\'aniel N\'ogr\'adi}$^{c}$,$\;$ 
B\'alint C. T\'oth$^{a}$,$\;$ 
K\'alm\'an K. Szab\'o$^{a}$,$\;$
Norbert Trombit\'as$^{c}$}

\author{\\
$^{a}$ Department of Physics, University of Wuppertal, Gau$\beta$strasse 20, D-42119, Germany\\
$^{b}$ J\"ulich Supercomputing Centre, Forschungszentrum J\"ulich, D-52425 J\"ulich, Germany\\
$^{c}$ Institute of Physics, E\"otv\"os University, P\'azm\'any P\'eter s\'et\'any 1/a, Budapest 1117, Hungary}

\author{\email{
borsanyi@uni-wuppertal.de, fodor@bodri.elte.hu, hch@physik.uni-wuppertal.de, 
katz@bodri.elte.hu, s.krieg@fz-juelich.de, nogradi@bodri.elte.hu, tothbalint@szofi.elte.hu, szaboka@general.elte.hu, 
trombitas@ludens.elte.hu}}

\abstract{
QCD is investigated at finite temperature using Wilson fermions in the fixed scale approach. A 2+1 flavor stout and clover improved action 
is used at four lattice spacings allowing for control over discretization errors. The light quark masses in this first study are
fixed to heavier than physical values.
The renormalized chiral condensate, quark number susceptibility and the Polyakov loop is measured and the results
are compared with the staggered formulation in the fixed $N_t$ approach. The Wilson results at the finest lattice spacing agree
with the staggered results at the highest $N_t$.
}

\FullConference{ The XXIX International Symposium on Lattice Field Theory - Lattice 2011\\
July 10-16, 2011\\
Squaw Valley, Lake Tahoe, California}

\begin{document}

\section{Introduction}
\label{introduction}


Extensive experimental work is currently being done with heavy ion collisions to study the QCD transition, most 
recently at the Relativistic Heavy Ion Collider, RHIC and at the Large Hadron Collider, LHC. Both for the 
cosmological transition and for RHIC/LHC, the net baryon densities are quite small, thus the baryonic chemical 
potentials ($\mu$) are much less than the typical hadron masses, $\mu$ is below 50~MeV at RHIC, even smaller at LHC 
and negligible in the early universe. Thus in this study we stay at $\mu=0$; for a review on thermodynamics see
\cite{Fodor:2009ax}.


When one analyzes the absolute scale or any other question related to the $T>0$ QCD transition for the physically 
relevant case two ingredients are quite important.

First of all, one should use physical quark masses. The nature of the transition strongly depends on the quark mass. 
Lattice studies and effective models showed that in the three flavor theory for small or large quark masses the 
transition is a first order phase transition, whereas for intermediate quark masses it is an analytic crossover. 


Secondly, the nature and other characteristics of the $T>0$ QCD transition is known to suffer from discretization 
errors \cite{deForcrand:2007rq,Endrodi:2007gc}. Let us mention one example which underlines the importance of 
removing these discretization effects by performing a controlled continuum extrapolation. The three flavor theory 
with a large, $a\approx 0.3$~fm lattice spacing and standard staggered action predicts a critical pseudoscalar mass 
of about 300~MeV \cite{Karsch:2001nf}. This point separates the first order and cross-over regions. If we took another 
discretization, with another discretization error, the critical pseudoscalar mass turns out to be much smaller, well 
below the physical pion mass of 135~MeV. The only way to determine the physical features of the transition is to 
carry out a careful continuum limit analysis. It can be safely done only in the so-called scaling regime. 
To carry out a controlled continuum extrapolation at least three lattice spacings in the scaling regime 
are needed, because two points will always lie on a 2-parameter curve describing the corrections to the continuum
results.

Numerically it is very demanding to fulfill both conditions. There are only a few cases for which this has been 
achieved. 
However it is important to note that fulfilling the second condition (at least 3 lattice spacings) 
without fulfilling the first one (physical quark masses) still leads to 
universal results. In other words continuum extrapolated results with non-physical quark masses are universal. 
These results are not the same as they are for physical 
quark masses, but they are well defined and unique. Contrary to this universality, fulfilling the first condition 
(physical quark mass) but not the second one (at least 3 lattice spacings) leads to non-universal, non-physical results. 
These results still have unknown discretization errors.

For this reason in this first study of QCD thermodynamics with Wilson fermions we have chosen to work at heavier than physical
quark masses but simulate at four lattice spacings in order to approach the continuum limit. 

In this paper we determine the temperature dependence of the chiral condensate, strange quark number
susceptibility and Polyakov-loop in 2+1 flavor QCD. We use Wilson fermions with six stout smearing and tree-level
clover improvement in the quark sector and tree-level improved fields in the gauge sector; for the details of the
action see \cite{Durr:2008rw}.

The structure of the paper is as follows. After this brief introduction 
the main features of the action and run parameters are listed in Section \ref{simulationpoints}. 
Our choice of renormalization procedures for the various measured
quantities are summarized in Section \ref{renormalization}. The results are given in Section \ref{results}. In Section
\ref{summary} we summarize and provide a short outlook. 

\section{Simulation points and techniques}
\label{simulationpoints}

The gauge action is the tree level improved Symanzik action \cite{Symanzik:1983dc}, the fermionic action is a 2+1 flavor stout
\cite{Morningstar:2003gk} and clover \cite{Sheikholeslami:1985ij} improved Wilson action.
The number of smearing steps is six and the smearing parameter is set at $0.11$; for the first large scale simulation 
employing stout smearing see \cite{Aoki:2005vt}. The clover improvement is tree level, i.e. $c_{sw} = 1$.  
In order to speed up the
molecular dynamics multiple time scales \cite{Sexton:1992nu}, Omelyan integrator \cite{Takaishi:2005tz} and even-odd preconditioning was used. For further
details on the algorithm see \cite{Durr:2008rw}
.
The finite temperature simulations were performed in the fixed scale approach \cite{Umeda:2008bd}. 

\begin{table}
\begin{center}
\begin{tabular}{|c|c|c|c|c|}
\hline
$\beta$ & $a m_{ud}$ & $a m_s$  & $N_s$ & $N_t$ \\
\hline
\hline
3.30 &  -0.0985  &   -0.0710 & 32 & 4 - 12, 32 \\
\hline
3.57 &  -0.0260  &   -0.0115 & 32 & 4 - 16, 64 \\
\hline
3.70 &  -0.0111  &   0.0     & 48 & 8 - 28, 48 \\
\hline
3.85 &  -0.00336 &   0.0050  & 64 & 12 - 28, 64 \\
\hline
\end{tabular}
\end{center}
\caption{Simulation parameters. The $N_t$ values used for the finite temperature runs and the values used
for the zero temperature runs are separated by a comma. For finite temperature only even $N_t$ values were used.}
\label{tab:parameters}
\end{table}

The simulation parameters are summarized in Table \ref{tab:parameters}. The simulations were performed at four lattice spacings
and dimensionless ratios of hadron masses were used to define the line of constant physics. The mass of the $\Omega$ baryon,
$m_{\Omega} = 1672\,MeV$, was used to set the scale. The ratio of the pion and omega mass was kept at heavier than physical value,
while the mostly light quark mass independent combination $(2m_K^2-m_\pi^2)/m_\Omega^2$ was kept close to its physical value.
More precisely the
combination which was tuned the most precisely is $(2m_K^2-m_\pi^2)/m_\pi^2$ which can also be used to define a physical
ratio $m_s/m_{ud}$ while $m_s$ itself was fixed by the requirement that as $m_{ud}$ is lowered to the physical point the kaon mass
becomes physical too. The large lattice volumes make sure finite volume effects are negligible, in all cases $m_\pi L > 7.9$. Table
\ref{tab:masses} summarizes the zero temperature hadron spectroscopy results. For more details see \cite{Durr:2008zz,Durr:2010vn,
Durr:2010aw}. 

\begin{table}
\begin{center}
\begin{tabular}{|c|c|c|c|c|c|}
\hline
$\beta$ & $m_\pi/m_\Omega$ & $m_K/m_\Omega$  & $(2m_K^2-m_\pi^2)/m_\pi^2 \approx m_s/m_{ud}$  & $a\; [fm]$ \\
\hline
\hline
3.30 &   0.336(5) &   0.378(6)  & 1.53(4)   &  0.137(2) \\
\hline
3.57 &   0.314(3) &   0.351(4)  & 1.50(2)  &   0.095(2) \\
\hline
3.70 &   0.317(2) &   0.356(3)  & 1.53(3)  &  0.0727(5) \\
\hline
3.85 &   0.322(6) &   0.362(6)  & 1.52(6)  &  0.0567(9) \\
\hline
\end{tabular}
\end{center}
\caption{Mass ratios from spectroscopy. The lattice spacings are given by $m_\Omega = 1672$ MeV.}
\label{tab:masses}
\end{table}

\section{Renormalization}
\label{renormalization}

Three quantities were measured, the chiral condensate ($\pbp$), the strange quark number susceptibility ($\chi_s$) and the
Polyakov loop ($P$). $\chi_s$ does not require renormalization but the others do. 

\subsection{Chiral condensate}

In this section the additive and multiplicative renormalization of the chiral condensate is summarized briefly, for more
details see \cite{Giusti:1998wy} which is based on \cite{Bochicchio:1985xa}.

The additive divergence in the bare condensate is $O(a^{-3})$ which at finite temperature can be removed by subtracting the zero
temperature value. The multiplicative divergence also needs to be removed by multiplying an appropriate expression of the quark
masses. This step is complicated by the fact that Wilson fermions break chiral symmetry explicitly but nevertheless the axial Ward
identity ensures that the following two combinations are finite and agree in the continuum limit \cite{Giusti:1998wy},
\bea
\label{final}
m_R \pbp_R(T) &=& m_{PCAC} Z_A \Delta_{\bar{\psi}\psi}(T) + O(a) \\
m_R \pbp_R(T) &=& 2 N_f m_{PCAC}^2 Z_A^2 \Delta_{PP}(T) + O(a)\nonumber
\eea
where $Z_A$ is a finite renormalization constant,
\bea
\label{diff1}
\Delta_{\bar{\psi}\psi}(T) &=& \pbpnull(T) - \pbpnull(T=0) \\
\Delta_{PP}(T) &=& \int d^4 x \langle P_0(x) P_0(0) \rangle(T) - \int d^4 x \langle P_0(x) P_0(0) \rangle(T=0) \nonumber
\eea
are the subtractions needed to cancel the additive divergences and $\pbpnull$ and $P_0(x)$ are the bare condensate and bare pseudo scalar 
density, respectively \cite{Giusti:1998wy}. In this first study we did not measure $Z_A$ but rather used the two expressions
(\ref{final}) to eliminate $Z_A$ (and also $m_{PCAC}$) and end up with the finite expression
\bea
\label{final2}
m_R \pbp_R(T) = \frac{ \Delta_{\bar{\psi} \psi}^2(T) }{ 2 N_f \Delta_{PP}(T) } + O(a)\;.
\eea

\subsection{Polyakov loop}

The divergence of the bare free energy means that the Polyakov loop also needs to be renormalized. A convenient choice of 
renormalization prescription is demanding a fixed
value $P_*$ for the renormalized Polyakov loop at a fixed but arbitrary temperature $T_* > T_c$. 
Then the renormalized Polyakov loop $P_R$ is given by
\bea
\label{pr2}
P_R(T) = \left( \frac{P_*}{P_0(T_*)} \right)^{\frac{T_*}{T}} P_0(T)
\eea
in terms of the bare Polyakov loop $P_0(T)$. We choose $T_* = 0.143 m_{\Omega}$ and $P_* = 1.2$. Other choices would simply
correspond to other renormalization schemes \cite{Aoki:2006br}.

\section{Results}
\label{results}

We have measured three quantities at four lattice spacing. The renormalized chiral condensate is 
sensitive to the chiral transition whereas the renormalized Polyakov loop and the
strange quark number susceptibility are sensitive to the remnant of the confinement-deconfinement transition.

Finite continuum results should agree between different discretizations provided the same renormalization conditions are used.
We have imposed the same conditions on staggered simulations hence these can be compared with the Wilson results; for a recent
review on staggered thermodynamics see \cite{Bazavov:2011nk}.

\begin{figure}
\begin{center}
\includegraphics[width=12cm]{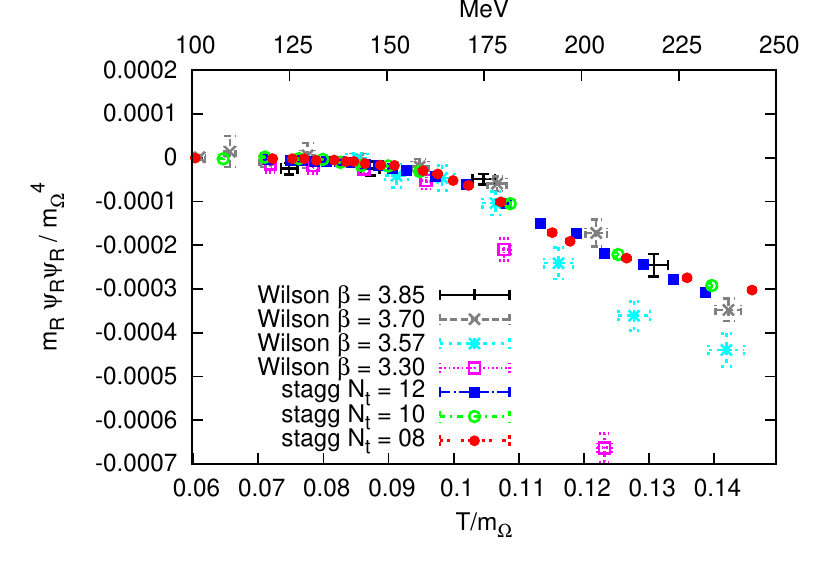}
\caption{Renormalized chiral condensate. The four lattice spacings with Wilson fermions (in the fixed scale approach) are compared with three lattice spacings
with staggered fermions in the fixed $N_t$ approach.\label{figpbp}}
\end{center}
\end{figure}

Figure \ref{figpbp} shows the renormalized chiral condensate at the four different lattice spacings together with staggered results
at three fixed $N_t$ values, 8, 10 and 12 (the staggered results were obtained using the fixed $N_t$ approach, for more details
see \cite{Aoki:2006we, Aoki:2006br,Aoki:2009sc,Borsanyi:2010bp}). As is apparent from the figure
the two finer lattice spacing values are small enough to observe only a small change in the chiral condensate at any temperature.
The rougest lattice spacing on the other hand is seen to include large discretization effects and is probably not in the scaling
region. What is also clear is that results at $N_t < 8$ are not trustworthy at any of the lattice spacings due to a too rough
resolution of the imaginary time direction (temperature). 
At only moderately high and small temperatures, particularly around $T_c$, the three finest lattice spacings are probably
in the scaling region and continuum extrapolation is possible.

\begin{figure}
\begin{center}
\includegraphics[width=12cm]{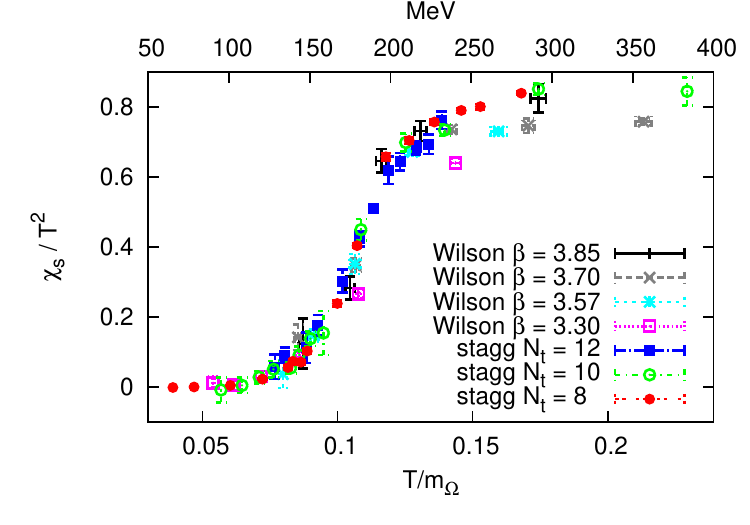}
\caption{Strange quark number susceptibility. The four lattice spacings with Wilson fermions (in the fixed scale approach) are compared with three lattice spacings
with staggered fermions in the fixed $N_t$ approach.\label{figqsusc}}
\end{center}
\end{figure}

The strange quark number susceptibility is a sum of two contributions, the connected and disconnected terms. The disconnected part
is a very noisy quantity (as usual) and a large number of random vectors are needed in order to evaluate it precisely
\cite{Ejiri:2009hq}.
The results for the four lattice spacings are shown on Figure \ref{figqsusc} together with the staggered results. 
We plot $\chi_s/T^2$ normalized by its Stefan-Boltzman limit value.

\begin{figure}
\begin{center}
\includegraphics[width=12cm]{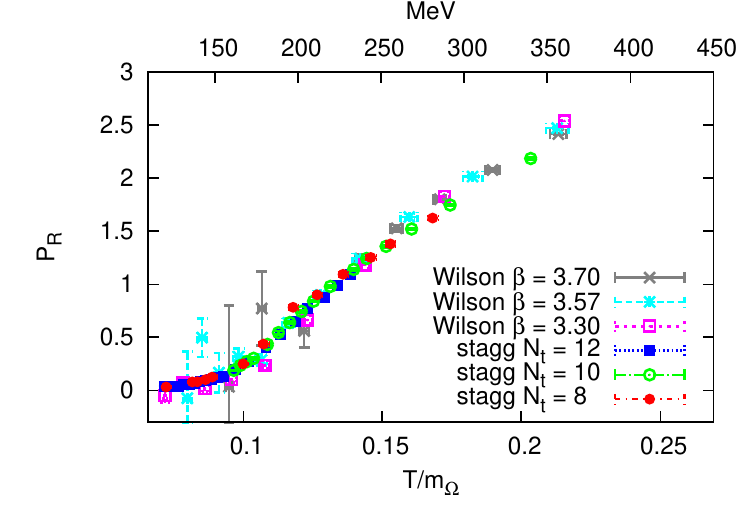}
\caption{Renormalized Polyakov loop. The four lattice spacings with Wilson fermions (in the fixed scale approach) are compared with three lattice spacings
with staggered fermions in the fixed $N_t$ approach.\label{figpolyakov}}
\end{center}
\end{figure}

The third quantity we measured is the renormalized Polyakov loop which is sensitive to the confinement-deconfinement transition
similarly to the quark number susceptibility. The results are shown on Figure \ref{figpolyakov}.

\section{Summary}
\label{summary}

In this work we have started our study of 2+1 flavor QCD thermodynamics with Wilson fermions. Special emphasis is placed on
obtaining results in the continuum limit, i.e. the simulations are performed at four lattice spacings. In this first study the
light quark masses were kept above the physical value but since continuum results are universal even at non-physical quark masses
a comparison is possible with continuum staggered results. We found nice agreement between the finest lattice spacings of the
Wilson and staggered formulations. In the future we plan to extend our results to lower pion masses and
also plan to use (\ref{final}) for $\pbp$ after measuring $Z_A$ and $m_{PCAC}$ because a priori it is not clear which
formula has the better scaling property.

\section*{Acknowledgment}

This work was supported by the DFG grants FO 502/1 and SFB-TR/55, and the EU Framework
Programme 7 grant (FP7/2007-2013)/ERC No 208740. For our calculations we used computer
time awarded by PRACE on Jugene in Juelich as well as GPU clusters in Budapest and
Wuppertal.

\end{document}